# Convolutional Neural Networks In Classifying Cancer Through DNA Methylation

Soham Chatterjee    Archana Iyer    Satya Avva    Abhai Kollara    Malaikannan Sankarasubbu

Saama Technologies AI Research Lab

Chennai, India

{s.chatterjee, a.iyer, satya.avva, a.kollara, malaikannan.sankarasubbu}@saama.com

*Abstract— DNA Methylation has been the most extensively studied epigenetic mark. Usually a change in the genotype, DNA sequence, leads to a change in the phenotype, observable characteristics of the individual. But DNA methylation, which happens in the context of CpG (cytosine and guanine bases linked by phosphate backbone) dinucleotides, does not lead to a change in the original DNA sequence, but has the potential to change the phenotype. DNA methylation is implicated in various biological processes and diseases including cancer. Hence there is a strong interest in understanding the DNA methylation patterns across various epigenetic related ailments in order to distinguish and diagnose the type of disease in its early stages. In this work, the relationship between methylated versus unmethylated CpG regions and cancer types is explored using Convolutional Neural Networks (CNNs). A CNN based Deep Learning model that can classify the cancer of a new DNA methylation profile based on the learning from publicly available DNA methylation datasets is then proposed.*

## I. Introduction

DNA Methylation is a chemical modification that refers to the addition of a Methyl group ( $CH_3$) via a covalent bond to the C-5 position of the Cytosine ring of DNA through a special class of enzymes called DNA Methyl Transferases (DNMTs) [1]. In humans, DNA Methylation occurs in the context of CpG dinucleotide sequence in somatic cells [2]. Exceptions to Methylation in context of CpG islands is observed in germline cells [3]. In human somatic cells most of the CpGs are Methylated. Unmethylated CpG regions are usually present at heads of the promoters in human genes, referred to as CpG islands [4][5].

Many techniques have been engineered over the past couple of decades to measure Methylated cytosines (5-mC). These techniques can vary from heat-based [6] to chemical-based methods [7][8]. Widely popular used technique to de-tect 5-mC is Bisulfite conversion technique used for preparing DNA for Methylation analysis on a gene-specific level[9]. In this technique unmethylated Cytosine is converted to Uracil while leaving 5-mC base intact to allow Methylation analysis at a single-nucleotide resolution.

DNA Methylation regulates many biological processes such as cell differentiation, development, cell reprogramming and proliferation through modulation of gene expression [10][11][12]. More studies on DNA Methylation revealed Differentially Methylated Regions (DMRs) that are DNA sequences that have varying Methylation profiles between different tissues, cell-lines, stages of development and organ-isms. Studying these regions are of immense interest in order to understand how differing Methylation profiles between biological samples can have an impact on implementing tissue-specific gene regulatory expression profiles [13].

Aberrant DMR patterns have been implicated in many diseases including autoimmune disorders [14], metabolic disorders [15], psychological disorders [16], aging [17] and cancer [18]. Therefore, quantifying the differences in DNA methylation across large numbers of samples and the identification of sample-specificity are critical steps in genomic function analysis, in normal and disease conditions.

Various studies on DNA methylation in the past have shown a strong link between differential DNA methylation and it's effects on regulating gene expression [19] [20]. Existing methods summarize differential methylation patterns localized for certain DNA regions which limits the scope of analysis. This study pushes the limits of analysis to a global scale through learning of changing DMR patterns and distinguishing them between different cancer types at a whole-genome level. This approach has immense potential to accurately diagnose the condition and even the stage of the disease. Based on this premise experiments were performed using high-throughput DNA methylation profiles from genome-wide analysis of various cancer types to classify them based on the learning of their DMR regions. This study utilized extensive DNA methylation data from The Cancer Genome Atlas (TCGA) repository [20], the details for which have been provided in the next section.

While there have been attempts at classifying diseases based on DNA methylation beta-values [21] [22], none have used a datasets and the choice of cutting-edge models as extensive and varied as the ones used in this work. The experiments carried out in this study involved training a Convolutional Neural Network (CNN) based Deep Learning classifier [23] on large volumes of publicly available DNA Methylation data obtained from high-throughput Bisulfite Sequencing methods, on cancer sub-types in order to create a representation of the differences between various cancer profiles.

## II. Data

For this study, cancer datasets were chosen due to the wealth of information available in The Cancer Genome Atlas

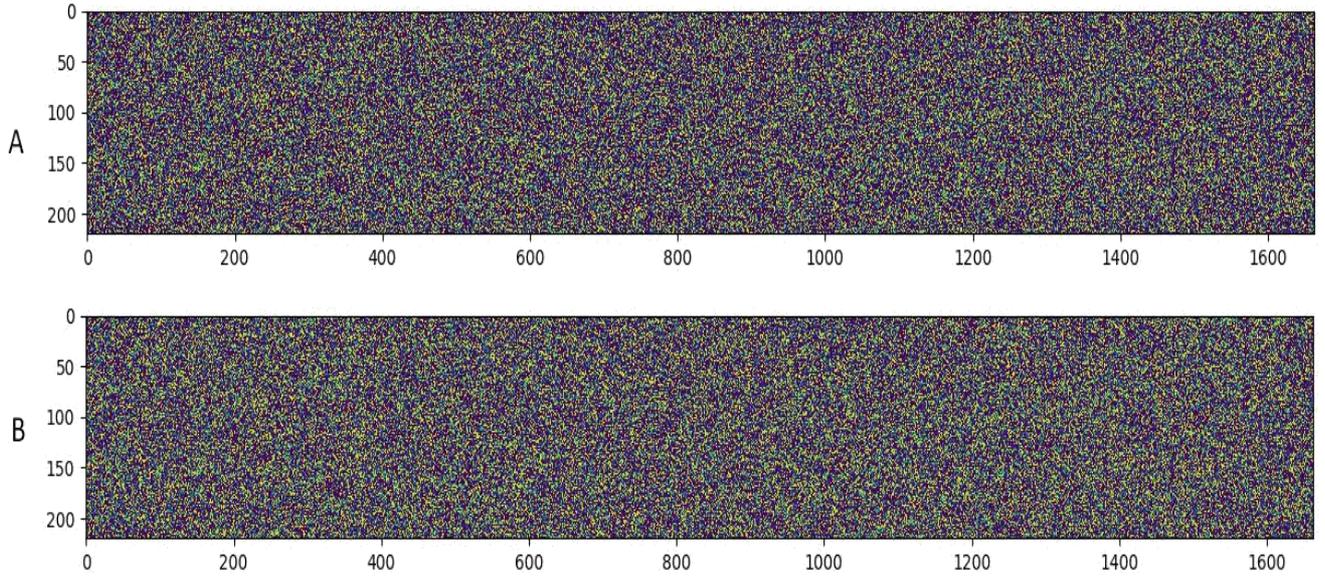

Fig. 1. Visualization of DNA methylation data of two different diseases transformed into images.
DNA Methylation profile data of individual samples, each transformed into an image of dimensions 220 x 1663. Each pixel value corresponds to the beta-value of a single CpG site. These images served as the input to train the proposed model.
A) Transformed image of BLCA - Urothelial Bladder Carcinoma
B) Transformed image of PRAD - Prostate Adenocarcinoma

(TCGA). The TCGA is a comprehensive archive of tumoral data containing the results of high-throughput experiments, mainly Next Generation Sequencing, for more than 30 cancer types [24]. TCGA contains 2.5 petabytes of whole-genome data that is made free available to the public and used widely by independent researchers contributing to thousands of studies on cancer. DNA methylation data was extracted using the TCGA2BED tool [25]. It contained DNA Methylation profiles from 33 different cancer types.

The data extracts from TCGA were samples that were processed to level 3 data type, which has the whole genome methylation calls for each CpG site, per sample in a tab-delimited file (.bed). All the DNA methylation profiles used in our experiment were aligned with the reference genome GRCh37 (hg19) [26]. About 10,000 DNA methylation pro-files encompassing 33 different cancer types were extracted from the TCGA. Data obtained from Illumina 450k DNA methylation samples were taken for our experiments. The detailed list is provided in Table I.

### III. DATA PREPROCESSING

The beta-value is a metric that corresponds to the ratio of intensities between the methylated probe and the overall intensity (sum of the methylated and unmethylated probes). This value ranges between 0 and 1, meaning that under ideal conditions the corresponding CpG region is either entirely unmethylated or totally methylated respectively [27].

Normalization is a technique used to scale input values to the neural network to a smaller range. This reduces training time and variance in the training loss graph. Since the Methylation values are already in between the values of 0 and 1, normalization is not done. Null values are replaced with zeros. While there are 33 different cancer diseases present in the dataset, training is not done with Ovarian Cancer (OV) DNA Methylation data as it was under represented (only 11 samples) in the dataset.

The data was divided into 3 parts: 20% of the data is used for testing and out of the remaining 80%, 10% is used as a validation set and the rest is used for training.

There are 365860 rows (each row corresponds to information on a CpG site) for each sample across all the cancer types. The beta-value of the corresponding row are extracted and converted into a matrix of shape (220; 1663). Each value in the matrix is defined by the beta-value (ranging from 0 to 1) of that CpG site. This matrix is given as an input image to the CNN model (described in the next section). Some of these images are presented in figure 1.

### IV. MODEL

Various CNN architectures were tested with different filter shapes that can properly find the relationship between the different DNA Methylation patterns that are thought to be correlated with diseases.

The model that is used is similar to that used by Yoon Kim for sentence classification [28]. The input to the model is an image,

$$x \in R^{hw} \qquad (1)$$

where h and w are the height and width of the image.

Convolutional operation with two filter shapes are used:

$$w_1 \in R^{kw} \qquad (2)$$

TABLE I
LIST OF CANCER TYPES AND NUMBER OF SAMPLES USED IN OUR STUDY

| Tumor Tag | Tumor Name | Experiment | # Aliquot | # Samples | # Patients |
|---|---|---|---|---|---|
| acc | Adrenocortical carcinoma | dnamethylation450 | 81 | 81 | 81 |
| blca | Bladder Urothelial Carcinoma | dnamethylation450 | 464 | 439 | 414 |
| brca | Breast Invasive carcinoma | dnamethylation450 | 930 | 897 | 794 |
| cesc | Cervical squamous cell carcinoma and endocervical adenocarcinoma | dnamethylation450 | 331 | 314 | 309 |
| chol | Cholangiocarcinoma | dnamethylation450 | 46 | 46 | 37 |
| coad | Colon adenocarcinoma | dnamethylation450 | 367 | 350 | 299 |
| dlbc | Lymphoid Neoplasm Diffuse Large B-cell Lymphoma | dnamethylation450 | 53 | 50 | 50 |
| esca | Esophageal carcinoma | dnamethylation450 | 212 | 204 | 187 |
| gbm | Glioblastoma multiforme | dnamethylation450 | 162 | 156 | 143 |
| hnsc | Head and Neck squamous cell carcinoma | dnamethylation450 | 598 | 582 | 530 |
| kich | Kidney Chromophobe | dnamethylation450 | 67 | 67 | 67 |
| kirc | Kidney renal clear cell carcinoma | dnamethylation450 | 496 | 485 | 321 |
| kirp | Kidney renal papillary cell carcinoma | dnamethylation450 | 336 | 323 | 277 |
| laml | Acute Myeloid Leukemia | dnamethylation450 | 194 | 194 | 194 |
| lgg | Brain Lower Grade Glioma | dnamethylation450 | 551 | 536 | 518 |
| lihc | Liver hepatocellular carcinoma | dnamethylation450 | 450 | 432 | 379 |
| luad | Lung adenocarcinoma | dnamethylation450 | 525 | 505 | 463 |
| lusc | Lung squamous cell carcinoma | dnamethylation450 | 413 | 408 | 371 |
| meso | Mesothelioma | dnamethylation450 | 89 | 88 | 88 |
| paad | Pancreatic adenocarcinoma | dnamethylation450 | 207 | 197 | 186 |
| pcpg | Pheochromocytoma and Paraganglioma | dnamethylation450 | 189 | 188 | 180 |
| prad | Prostate adenocarcinoma | dnamethylation450 | 573 | 555 | 500 |
| read | Rectum adenocarcinoma | dnamethylation450 | 113 | 108 | 100 |
| sarc | Sarcoma | dnamethylation450 | 285 | 271 | 263 |
| skcm | Skin Cutaneous Melanoma | dnamethylation450 | 493 | 478 | 473 |
| stad | Stomach adenocarcinoma | dnamethylation450 | 413 | 400 | 398 |
| tgct | Testicular Germ Cell Tumors | dnamethylation450 | 158 | 157 | 151 |
| thca | Thyroid carcinoma | dnamethylation450 | 588 | 573 | 509 |
| thym | Thymoma | dnamethylation450 | 128 | 127 | 125 |
| ucec | Uterine Corpus Endometrial Carcinoma | dnamethylation450 | 508 | 484 | 446 |
| ucs | Uterine Carcinosarcoma | dnamethylation450 | 58 | 58 | 58 |
| uvm | Uveal Melanoma | dnamethylation450 | 81 | 81 | 81 |
| | | Total | 10,159 | 9,834 | 8,992 |

$$w_2 \in R^{hk} \qquad (3)$$

where k is the window size used to produce a new feature. The filters are applied successively to get an output feature map for that filter. While filter $w_1$ is applied along the height of the image, filter $w_2$ searches for features along the width of the image. Unlike the model proposed by Yoon Kim in [28], a max pooling operation is not performed on the feature maps obtained. Instead all the features of the output feature map are retained.

The model has a single filter layer with multiple such filters with variable window sizes to extract numerous features. The filters are grouped into four types with five filters in each type based on the window size of the filter.

While all the $w_1$ convolutional filters have the same width (equal to the width of the image), their heights are 1, 3, 5 and 7. Whereas all the $w_2$ convolutional filters have the same height (equal to the height of the image) and their widths are 1, 3, 5 or 7.

The ideology behind such a filter shape is to allow the narrow filter to learn the smaller or local features whereas the wider filters will be able to learn larger features that are more spread out over different DNA methylation regions.

After the filter layer, all the features are concatenated. This concatenated output is then fed into two fully connected layers. These layers have 512 and 128 neurons followed by a softmax layer with 32 neurons for each of the classes. The model architecture has been illustrated in figure 2.

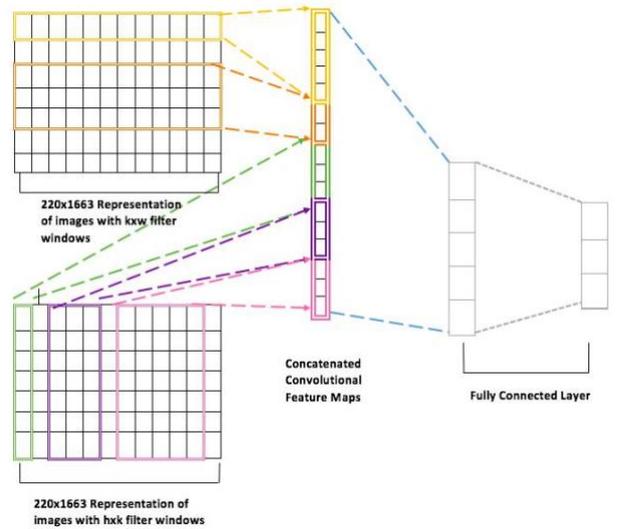

Fig. 2. Model Architecture

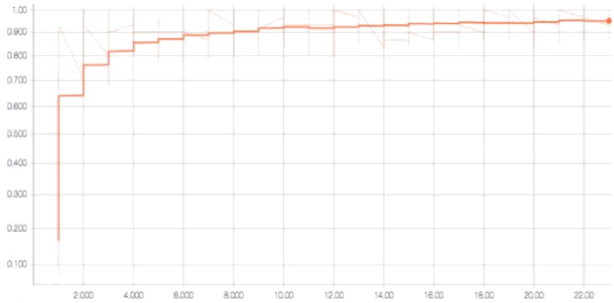

Fig. 3. Accuracy vs Epochs

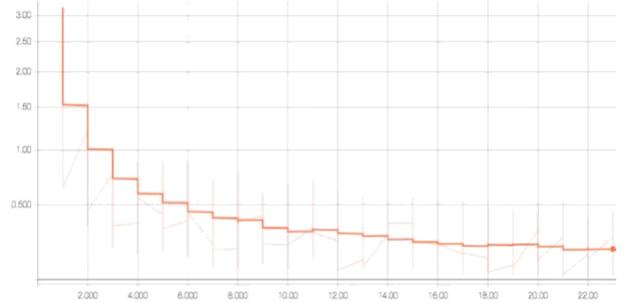

Fig. 4. Loss vs Epochs

A. *Model Training*

Training was done using mini batch stochastic gradient descent (SGD) using shuffled mini batches of size 30. Adam optimizer with an initial learning rate of 1e- 4 was used [29]. All our parameters were initialized with zero-centered Normal distribution with standard deviation 0.02. Training was halted if the model accuracy on the validation set decreased twice by more than 0.5%. All of the trained models (discussed in the next section) converged before 30 epochs.

B. *Experiment*

While the model described above gave us the best results, other models were also trained. These models had different hyper-parameters and filter shapes. However, the training contexts remained the same for all the models.

Initially a baseline CNN with 3 filter layers was trained with 10 filters in each layer followed by a similar fully connected layer (CNN_Base) to get a baseline result. This gave a train accuracy of 89.38%. The kernels used had shapes 3x3. Increase in the number of layers or filters had no significant improvements in performance.

This could be because of the shape of the filters used. Filters with a square shape are not able to capture the larger and more spread out relationships between DMRs. Moreover, DNA Methylated regions are spread out in the horizontal direction (in our image) and a square filter shape may not be able to capture those relationships properly.

Since cancer like diseases are usually not caused by a single gene but a subset of genes which could be located far apart on the genome, DMRs responsible for regulating such genes would also be far apart. Square filters would be efficient in capturing changing DMRs when they are located nearby, but may not be good in capturing changing DMRs when they are distantly located. Hence rectangular filters that can capture changing DMR patterns placed near and slightly far were used.

To test our hypothesis, a rectangular filter was used that spanned the entire width of the image with a height of 1 row. This gave an improved accuracy. Using a combination of filter window sizes of 1, 3, 5 and 7 further improved the accuracy (CNN_Width). As discussed previously, this is because of how the CpG sites are widely distributed across the genome.

For further improvements in accuracy height-wise convolutional filters was also added that should take into account the smaller features in the Methylation patterns that are spread out over the entire DNA data. This is the final model.(CNN_Height_Width).

TABLE II
COMPARISON OF ACCURACY ON DIFFERENT PARAMETERS OF ARCHITECTURE

| Model | Train Accuracy | Test Accuracy |
|---|---|---|
| CNN_Base | 89.38% | 84.34% |
| CNN_Width | 95.91% | 91.97% |
| CNN_Height | 95.71% | 92.06% |
| CNN_Height_Width | 96.54% | 92.87% |

To study the effects of the height-wise filters, a model with only height-wise convolutional filters (CNN_Height) was trained. The result of this shows that there are many features that are spread-out in the Methylation data that have a correlation with the disease.

The training results of all the models are tabulated in Table II.

V. RESULTS AND DISCUSSION

The aim was to build a model that could learn the minute differences of changing DMR patterns on a genome-wide scale and generalize well across various cancer types to accurately classify the cancer type. The DNA methylation data was evaluated for 32 cancer types in our experiment. About 10,000 samples was examined that represent each of the cancer types to train our model.

A 'black-box' approach was taken where the analysis was performed on a global scale across entire genomes as opposed to targeting specific DNA regions and cancer type. This approach allowed the models to capture the changing DMR landscape among various cancer types from whole-genome DNA methylation datasets. This enabled the model to learn the specific DMR patterns within a disease and apply its learning in predicting the cancer type from test data that the model has not seen before.

The proposed model achieves a training accuracy of 96.54% and a test accuracy of 92.87%. The accuracy and loss graphs for the same have been plotted in figure 3 and figure 4.

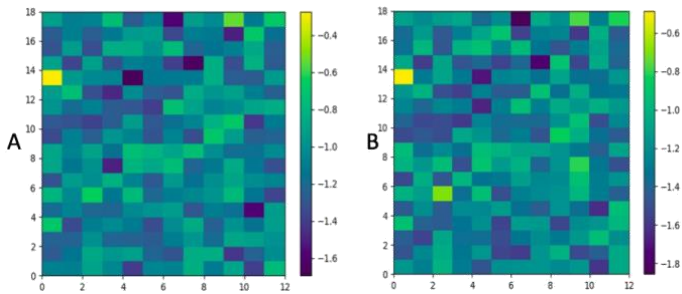

Fig. 5. Importance scoring of genomic features between two diseases. A) BLCA and B) PRAD

Further analysis was performed to retrieve the relative importance of the ground-level features, genomic regions that are collection of CpG sites, which formed the basis for classification of diseases in our models. Importance scoring of features in the context of this study allows for realizing highly influential genomic regions with DMR patterns associated with the diseases. Although it is not possible to identify the exact CpG sites that are highly influential, since the filters look at a large number of genomic sites, importance scoring provides an overview of ability of the model in this study in efficiently capturing the changing DMR landscape across various diseases from the input.

Importance scoring of the features comparing two diseases (BCLA and PRAD) have been presented in figure 5 as a heat-map. Scale value for the heat-map is provided to understand the relative importance. The scale indicates high to low (top to down) importance of features retrieved on a window size 5 for both the diseases. Comparing heat-maps in figure 5 reveals that albeit very few differences in the important genomic features between the two diseases the model is efficiently able to learn minute changes in the genomic landscape and apply its learning to differentiate between the diseases. Generating more verbose visualizations can give us a better understanding of which CpG sites may correlate with a certain disease and will be done in a future work.

## VI. CONCLUSION

Traditional methods to detect cancer are not very efficient. They require a lot of manual effort, from sample preparation to detecting the disease. Many of these techniques have lower accuracy and depend on bio-markers that usually become prevalent in the later stages of the disease. Over-diagnosis is also an on-going issue whereby false-positives leads to further investigation and adds anxiety and financial burden on patients. Hence there is urgent requirement to greatly improve the state of cancer screening.

In comparison to other biochemical techniques, analysis of DNA methylation profiles on a whole-genome scale holds significant promise in diagnosing cancer with much better precision. However, handling large datasets can create roadblocks in efficiently analyzing and classifying disease types as the existing methods are drastically slow.

Utilizing deep learning methods gives a major boost to this analysis and puts it at an advantage over other methods as deep learning methods are not only tailored for processing big datasets but also for providing accurate results.

In this work a CNN based classifier was used to classify diseases based on identification of relationship between DNA Methylation of specific CpG sites and diseases. In the experiment a model with both height-wise and width-wise convolutional filters was trained and seen to give the best accuracy.

The time taken to obtain the results for a batch of 32 test samples is 0.35 seconds, whereas the time taken to classify the methylation profile for a single test sample is 0.26 seconds (since batching improves TensorFlow performance). These values are those obtained using TensorFlow run on a system with an Intel i7 processor with 6 cores and 12 MB of RAM.

In conclusion the proposed method in this paper can be used in classification of cancer type using DNA methylation information on a genome-wide scale with high accuracy. Identifying and targeting the CpG sites that are responsible for giving rise to the condition within each disease could make our approach more powerful and achieve top accuracy levels. Finally, we present our method to be advanced and highly efficient that can be applicable to analyze other epi-genetic related diseases as well, with additional parameters for input as required.

## ACKNOWLEDGMENT

The authors would like to thank Dr. Anand Dubey, Mr. Nikhil Gopinath and Professor Aravindan Chandrabose for reviewing the manuscript and for providing their technical inputs.